\documentclass{article}

\usepackage{spconf}

\usepackage{url}            %
\usepackage{booktabs}       %
\usepackage{amsfonts}       %
\usepackage{nicefrac}       %
\usepackage{xcolor}         %
\usepackage{multirow}

\usepackage{spconf,amsmath,amssymb,bm,graphicx}
\usepackage[hidelinks]{hyperref}
\usepackage[capitalize,nameinlink]{cleveref}
\usepackage{subcaption}
\usepackage{booktabs}
\usepackage[export]{adjustbox}
\usepackage{microtype}
\usepackage{flushend}
\usepackage[backend=biber, bibencoding=utf8,
            style=ieee,
            maxbibnames=3,
            doi=false,
            url=false,
            isbn=false, natbib, maxcitenames=1, mincitenames=1]{biblatex}
\AtEveryBibitem{%
  \ifentrytype{inproceedings}{%
    \clearfield{pages}%
    \clearlist{publisher}%
    \clearlist{organization}%
    \clearlist{location}}{}%
  \ifentrytype{misc}{%
    \clearfield{eprintclass}{} }}

\usepackage[subtle]{savetrees}

\title{LipDiffuser: Lip-to-Speech Generation with Conditional Diffusion Models}

\name{Julius Richter$^*$, Danilo {de Oliveira}$^*$, Tal Peer, Timo Gerkmann}
\address{Signal Processing, University of Hamburg, Germany}

\usepackage[nolist]{acronym}

\begin{acronym}
\acro{ASR}{automatic speech recognition}
\acro{DNN}{deep neural network}
\acro{WER}{word error rate}
\acro{SNR}{signal-to-noise ratio}
\acro{STFT}{short-time Fourier transform}
\acro{LSE-C}{lip sync error confidence}
\acro{LSE-D}{lip sync error distance}
\acro{PESQ}{perceptual evaluation of speech quality}
\acro{ESTOI}{extended short-time objective intelligibility}
\acro{DNSMOS}{deep noise suppression mean opinion score}
\acro{LPS}{Levenshtein phoneme similarity}
\acro{CTC}{connectionist temporal classification}
\acro{MP-ADM}{magnitude-preserving ablated diffusion model}
\acro{ADM}{ablated diffusion model}
\acro{SDE}{stochastic differential equation}
\acro{SSL}{self-supervised learning}
\acro{MP-FiLM}{magnitude-preserving feature-wise linear modulation}
\acro{FiLM}{feature-wise linear modulation}
\acro{EEG}{electroencephalography}
\acro{FFT}{fast Fourier transform}
\acro{GPU}{graphics processing unit}
\acro{ROI}{region-of-interest}
\acro{EMA}{exponenibug.face_alignment import FANPredictortial moving average}
\acro{DAC}{descript audio codec}
\acro{VSR}{visual speech recognition}
\acro{EMA}{exponential moving average}
\acro{SE}{speech enhancement}
\acro{AVSE}{audio-visual speech enhancement}
\end{acronym}

\begin{document}
\ninept

\maketitle

\renewcommand{\thefootnote}{\fnsymbol{footnote}} 
\footnotetext[1]{Authors contributed equally to this work.} 
\renewcommand{\thefootnote}{\arabic{footnote}} 

\begin{abstract}
We present \emph{LipDiffuser}, a conditional diffusion model for lip-to-speech generation synthesizing natural and intelligible speech directly from silent video recordings.
Our approach leverages the magnitude-preserving ablated diffusion model (MP-ADM) architecture as a denoiser model. 
To effectively condition the model, we incorporate visual features using magnitude-preserving feature-wise linear modulation (MP-FiLM) alongside speaker embeddings. 
A neural vocoder then reconstructs the speech waveform from the generated mel-spectrograms. 
Evaluations on \emph{LRS3} demonstrate that \emph{LipDiffuser} outperforms existing lip-to-speech baselines in perceptual speech quality and speaker similarity, while remaining competitive in downstream automatic speech recognition. These findings are also supported by a formal listening experiment.
\end{abstract}


\section{Introduction}\vspace{-5px}

Real-world video recordings often suffer from a very low-quality audio track. This can result from a noisy recording environment, inadequate equipment, or storage media corruption.
In severe cases, the audio track may be completely unusable or even missing. 
In the context of human speech, visual and auditory information both play a role in the ability to understand what a person is saying~\cite{mcgurk1976hearing}. 
However, these two modalities are not equally important; while missing visual information can lead to reduced intelligibility, the lack of audio is practically destructive to speech communication~\cite{dodd1977role}, except for individuals highly trained in lip reading~\cite{dodd1987hearing}.
Distorted audio can be restored using \ac{SE} systems, which may also harness the visual modality~\cite{girin1995noisy,richter2023audio}, but these systems have their limits, and they tend to perform poorly in adverse noise conditions. 

Lip-to-speech refers to the process of generating natural-sounding speech from a silent video of a person speaking. 
Such systems enable speech understanding even when the audio track is missing or heavily distorted, including cases with negative \acp{SNR}.
A high-quality lip-to-speech system should produce speech that is intelligible, synchronized with the speaker’s lip movements, and perceptually natural.
Ideally, the speech signal must also match the speaker's characteristics, including age, gender, and accent.
A major challenge for lip-to-speech techniques stems from the one-to-many mapping between visemes and phonemes~\cite[Table 1]{richter2022continuous}; different phonemes can correspond to the same sequence of lip movements, leading to ambiguity. 
This issue can be alleviated by including a longer temporal context in the generation process~\cite{martinez2020lipreading}. 
Another difficulty is alignment and synchronization of the visual cues from the lip movements with the generated audio, considering the large differences in feature rates between the visual and audio signals (e.g., 25$\,$Hz for video compared to 48$\,$kHz for audio).

In this work, we introduce \emph{LipDiffuser}, a lip-to-speech method based on a conditional diffusion model. 
The proposed method comprises a video encoder, a speaker encoder, a denoiser model, and a neural vocoder. 
For the denoiser model, we utilize the \ac{MP-ADM} architecture~\cite{karras2024analyzing}, which builds upon the \ac{ADM} architecture~\cite{dhariwal2021diffusion} by introducing a series of modifications that significantly enhance output quality while retaining the overall structure.
During diffusion sampling, the denoiser model takes as input the current process state, video features extracted by the video encoder, the current diffusion timestep, and a speaker embedding from the speaker encoder, and predicts a mel-spectrogram.
To effectively incorporate visual cues, we propose integrating video features matched via interpolation and downsampling into the denoising network’s decoder using our proposed \ac{MP-FiLM} layers.
After the reverse diffusion process, the neural vocoder synthesizes the final audio waveform from the generated mel-spectrogram.

We conduct experiments on \emph{LRS3}~\cite{afouras2018lrs3}. 
To enable comparison with \ac{AVSE} methods, we create the \emph{LRS3-CHiME3} dataset by mixing speech from \emph{LRS3} with noise files from \emph{CHiME3}~\cite{barker2015third}. 
Our findings suggest that lip-to-speech approaches can offer advantages over \ac{AVSE}, particularly when the input audio is severely degraded (\ac{SNR}  $<-5\,$dB). Specifically, we show that our proposed \emph{LipDiffuser} model outperforms all lip-to-speech baselines in terms of speech quality and speaker similarity, while remaining competitive in \ac{ASR} performance and synchrony. Besides instrumental metrics, results on speech quality and speaker similarity are also supported by formal listening experiments.
Furthermore, cross-dataset evaluations demonstrate the strong generalization capabilities of our method in both speech quality and speaker similarity. 
Audio-visual examples are available online.\footnote{\url{https://sp-uhh.github.io/lip2speech}} Code and pretrained checkpoints will be made available upon acceptance.

\section{Related work}\vspace{-5px}

\subsection{Lip-to-speech}\vspace{-3px}

A range of learning-based lip-to-speech methods has been introduced in recent years. 
These approaches vary regarding the underlying model architectures and learning paradigms, the types of visual and additional information leveraged, and the strategies used to guide the process during training and inference.

\citet{kim2023lip} present \emph{Lip2Speech}, a lip-to-speech method that includes a visual front-end, a \emph{Conformer} model for capturing temporal relationships, and a mel-spectrogram generator conditioned on speaker embeddings.
They propose a multi-task learning approach that consists of: (a) a \ac{CTC} loss between the visual representations and the text targets; (b) text prediction utilizing a pretrained \ac{ASR} model; and (c) a mel-spectrogram reconstruction loss. 
Finally, they use the Griffin-Lim algorithm to convert the mel-spectrogram into a waveform.

\citet{choi2023intelligible} propose a lip-to-speech model that employs speech representations from an \ac{SSL} model.
In a multi-task learning framework, their model predicts both mel-spectrograms and speech representations derived from a pretrained \ac{SSL} model.
In addition, they develop a vocoder that generates the waveform with the mel-spectrogram and speech units. 
During training, they augment the input mel-spectrograms with blur and noise to help the vocoder learn to generate waveforms from the generated mel-spectrograms by referencing the speech units.
In a follow-up work, \citet{choi2023diffv2s} present \emph{DiffV2S}, a diffusion-based lip-to-speech method based on a speaker embedding derived from the visual input. 
This approach addresses the challenge of inferring the speaker's characteristics without audio.

\citet{yemini2024lipvoicer} introduce \emph{LipVoicer}, a lip-to-speech method that uses a diffusion model conditioned on lip video and incorporates a classifier-guidance mechanism based on text. 
A pretrained \ac{ASR} model acts as the classifier, whereas a pretrained lip-reading network predicts the spoken text from the silent video, offering guidance for the score model.
They train the diffusion model to generate mel-spectrograms and utilize a neural vocoder to produce the raw audio.

In contrast to the above-mentioned methods, our \emph{LipDiffuser} model achieves aligned conditioning by interpolating the video features to match the temporal resolution of the audio features within the denoiser model, and fusing them through \ac{MP-FiLM} layers.
In addition, we pre-enhance the audio track of the ``in-the-wild'' audio-visual training data to increase the audio generation quality of our model, and pretrain the model with audio-only data first.

\subsection{Audio-only and audio-visual speech enhancement}\vspace{-3px}

\ac{SE} refers to the process of improving the quality and intelligibility of a speech signal by reducing noise and reverberation~\cite{vincent2018audio}. 
Numerous computational methods have been developed for this task, most operating in the time-frequency domain using the \ac{STFT}. 
Machine learning-based \ac{SE} methods are commonly divided into predictive and generative learning paradigms~\cite{lemercier2025diffusion}, and they vary in the types of information leveraged to estimate the clean speech signal.
While practical approaches often exploit multi-channel input from microphone arrays to utilize spatial information for improved performance in adverse acoustic environments, the present work focuses on single-channel \ac{SE}, relying solely on audio captured from a single microphone.

Human perception naturally integrates auditory and visual information to enhance speech understanding, particularly in challenging listening environments~\cite{stein1993merging}. 
Neuroimaging and behavioral studies have shown that visual cues, such as a speaker’s lip movements, can significantly improve both speech intelligibility and localization~\cite{calvert1997activation}. This inherent crossmodal integration motivates the development of \ac{AVSE} methods, aiming to replicate these perceptual advantages in machine learning systems~\cite{girin1995noisy}. By leveraging both audio and visual modalities, modern systems seek to enhance speech quality and intelligibility more robustly than is possible with audio alone~\cite{richter2023audio}.

\section{Method}\vspace{-5px}

\emph{LipDiffuser} consists of a video encoder $E_v$, a speaker encoder $E_s$, a denoiser model $D_\theta$ parameterized by $\theta$, and a neural vocoder $D_a$.
The denoiser model receives as inputs the process state $\mathbf x_t \in \mathbb R^{n_a}$, a speaker embedding $\mathbf s \in \mathbb R^{n_s}$ from the speaker encoder, video features $\mathbf v \in \mathbb R^{n_v}$ from the video encoder, and the process time $t$, and predicts the mel-spectrogram $\hat{\mathbf x} \in \mathbb R^{n_a}$.

\subsection{Training objective}\vspace{-3px}
We use denoising score matching as our training objective. 
For a process time $t$, the training objective is defined as minimizing 
\begin{equation}
    \mathcal J (D_\theta, t) = \mathbb E_{\mathbf x,\mathbf n, \mathbf s, \mathbf v} \left[\| D_\theta(\mathbf x + \mathbf n, \mathbf s, \mathbf v, t) - \mathbf x\|_2^2\right],
\end{equation}
where $(\mathbf x, \mathbf s, \mathbf v)\!\sim\!p_\text{data}(\mathbf x, \mathbf s, \mathbf v)$ is sampled from the dataset, and $\mathbf n\!\sim\!\mathcal N(\mathbf 0, \sigma(t) \mathbf I)$ is a random Gaussian vector with time-dependent standard deviation $\sigma(t)=t$.
The overall training objective is defined as a weighted expectation of $\mathcal J (D_\theta, t)$ over the process time, 
\begin{equation} 
    \mathcal J(\theta') = \mathbb E_t \left[\frac{\lambda(t)}{\exp{u_{\tilde{\theta}}(t)}} \mathcal J (D_\theta; t)+ u_{\tilde{\theta}}(t)\right],
\end{equation}
where $\ln(t)\!\sim \mathcal{N}(P_{mean},P_{std})$ with hyperparameters $P_{mean}$ and $P_{std}$, and $\lambda: \mathbb R \to \mathbb R$ is a time-dependent weight.
Following uncertainty-based multi-task learning~\cite{kendall2018multi}, the loss is further weighted by $u_{\tilde{\theta}}(t)$, a learnable linear projection of the process time parameterized by $\tilde{\theta}$, representing the uncertainty of the model. 
At the same time, the model is penalized for this uncertainty, encouraging $u_{\tilde{\theta}}(t)$ to be as low as possible. 
The full parameter set $\theta' = \{\theta, \tilde{\theta}\}$ thus includes both the denoiser model parameters and those of the learnable linear projection.

Following \citet{karras2022elucidating}, we define the denoiser model $D_\theta$ as 
\begin{equation}
    D_\theta(\mathbf x_t, \mathbf s, \mathbf v, t) = c_\text{skip}(t) \mathbf x_t + c_\text{out}(t) F_\theta (c_\text{in}(t) \mathbf x_t, \mathbf s, \mathbf v, t),
    \label{eq:preconditioning}
\end{equation}
where $c_\text{skip}: \mathbb R \to \mathbb R$ is a skip scaling controlling the skip connection of $\mathbf x_t$, $c_\text{out}: \mathbb R \to \mathbb R$ is an output scaling, and $c_\text{in}: \mathbb R \to \mathbb R$ is an input scaling. 
The function $F_\theta: \mathbb R^{n_a}\!\times\!\mathbb R^{n_s}\!\times\!\mathbb R^{n_v}\!\times\!\mathbb R\rightarrow \mathbb R^{n_a}$ is a neural network parameterized by $\theta$.

\subsection{Network architecture}\vspace{-3px}

We utilize the \ac{MP-ADM} architecture for our neural network~\cite{karras2024analyzing}. 
This architecture builds on the \ac{ADM} architecture~\cite{dhariwal2021diffusion}, introducing a series of modifications that retain the overall structure while significantly enhancing output quality. 
In particular, the network layers are carefully designed to preserve the expected magnitudes of activations and weight matrices during training. 
To condition the network on video features, we integrate \ac{FiLM}~\cite{perez2018film} into the \ac{MP-ADM} architecture. \ac{FiLM} has been successfully applied in conditioning audio source separation~\cite{liu2022separate} and \ac{SE}~\cite{deoliveira2024laser} systems on language prompts, and conditioning audio on video features for audio-visual sound event recognition \cite{brousmiche2022multimodal}.
Here, we introduce \ac{MP-FiLM}, which we motivate as follows.

\subsection{Magnitude-preserving FiLM}\vspace{-3px}
For an input $\mathbf x\in \mathbb R^{d_x}$ and conditioning variable $\mathbf c \in \mathbb R^{d_c}$, \ac{FiLM} layers are channel-wise defined as
\begin{equation}\label{eq:orig_film}
    \operatorname{FiLM}(\mathbf{x},\mathbf c | \theta, \phi) = \gamma_\theta(\mathbf c)\odot\mathbf{x} + \beta_\phi(\mathbf c), 
\end{equation}
where $\gamma_\theta\!:\!\mathbb R^{d_c}\!\rightarrow\!\mathbb R^{d_x}$ and $\beta_\phi\!:\!\mathbb R^{d_c}\!\rightarrow\!\mathbb R^{d_x}$ are neural networks parameterized by $\theta$ and $\phi$, and $\odot$ represents the Hadamard product~\cite{perez2018film}.
To ensure magnitude preservation of the addition in \eqref{eq:orig_film}, we define \ac{MP-FiLM} as
\begin{equation}
    \operatorname{MP-FiLM}(\mathbf{x},\mathbf c | \theta, \phi) = \frac{(\mathbf 1-\gamma_\theta(\mathbf c))\odot\mathbf{x}+\gamma_\theta(\mathbf c) \odot\beta_\phi(\mathbf c)}{\sqrt{(\mathbf 1-\gamma_\theta(\mathbf c))^2+\gamma_\theta(\mathbf c)^2}}.
\end{equation}

Unlike the original \ac{FiLM}, here the scaling and shifting of the network's activations are conducted per frame and not globally, so that each lip movement conditions the corresponding audio frame. With that intent, video features are initially interpolated in the time dimension to match the audio frame rate. We implement $\beta_\phi$ and $\gamma_\theta$ with two-layered convolutional blocks, containing a convolutional layer of kernel size 5, to account for potential misalignments between audio and lip movement, followed by a pointwise convolution. 
The neural network $\gamma_\theta$ contains an additional learned gain initialized at 0, followed by a clamping operation to bound the output between 0 and 1, defining the relative contribution of the video features at each fusion stage.

\subsection{Inference}\vspace{-3px}

For generating speech from video, we first extract video features $\mathbf{v} = E_v (\tilde{\mathbf{v}})$ from preprocessed video frames $\tilde{\mathbf{v}} \in \mathbb{R}^{N \times H \times W}$ using the video encoder $E_v$, where $\tilde{\mathbf{v}}$ is a grayscale video with $N$ frames, height $H$, and width $W$ depicting the speaker's lip movements.
We extract a speaker embedding $\mathbf{s} = E_s(\tilde{\mathbf s})$ using the speaker encoder $E_s$, where $\tilde{\mathbf s} \in \mathbb{R}^{n_{\tilde{\mathbf s}}}$ is enrollment data of the target speaker's voice containing $n_{\tilde{\mathbf s}}$ samples.
Given the precomputed video features and the speaker embedding, we perform reverse diffusion using the trained denoiser model $D_\theta$.
We use the second-order deterministic sampler from \cite{karras2022elucidating} with $M=32$ sampling steps to generate a mel-spectrogram $\hat{\mathbf x} \in \mathbb R^{n_a}$.
Finally, we synthesize the audio signal $\tilde{\mathbf{x}} = D_a(\hat{\mathbf{x}})$ using the neural vocoder $D_a$, where $\tilde{\mathbf{x}} \in \mathbb{R}^{n_{\tilde{x}}}$ denotes the estimated time-domain speech signal containing $n_{\tilde{x}}$ samples.

\section{Experimental setup}\vspace{-5px}

We train our models for 32M samples using batch size 256, reference (peak) learning rate of 0.005, with linear ramp up followed by inverse square root decay, following \cite{karras2024analyzing}. The optimizer is ADAM~\cite{kingma2015adam}, with $\beta_1=0.9$ and $\beta_2=0.99$. The weighting function is $\lambda(t)=(\sigma(t)^2+\sigma_{data}^2)/(\sigma(t)\,\sigma_{data})^2$, where $\sigma_{data}$ is calculated on a subset of the training set. For the noise level distribution, we set $P_{mean}=-1.2$ and $P_{std}=1.2$, following \cite{karras2024analyzing}. The denoiser model contains $\sim205$M parameters. Training is conducted on NVIDIA H100 \acp{GPU}, typically requiring 60 \ac{GPU} hours.

\subsection{Data}\vspace{-3px}
\label{sec:data}
We utilize the \emph{LRS3} dataset~\cite{afouras2018lrs3} for training, validation, and the test.
\emph{LRS3} is an ``in-the-wild'' dataset comprising audio-visual recordings from TED Talks, in which the audio track often includes background noise and reverberation.
Therefore, we use a pretrained generative \ac{SE} model, specifically the Schrödinger bridge~\cite{jukic2024schr, richter2025investigating}, to enhance the audio tracks of the \emph{LRS3} dataset. 
The preprocessing model has been trained on the \emph{EARS-WHAM}~\cite{richter2024ears} and \emph{VB-DMD}~\cite{valentini2016investigating} datasets, for which a checkpoint is available online.\footnote{\url{https://github.com/sp-uhh/sgmse}}
For the method to work, we have found that an audio-only pretraining stage is required, without video conditioning features. To increase the variety and quantity of training data, we pretrain the model on the 960 hours of the \emph{LibriSpeech} dataset \cite{panayotov2015librispeech}, also preprocessed by the \ac{SE} model.

We view lip-to-speech as an extreme case of \ac{AVSE}, where the audio is either completely absent or heavily corrupted. 
Consequently, we create an \ac{AVSE} benchmark called \emph{LRS3-CHiME3}. 
This benchmark includes six subsets, each with a distinct \ac{SNR} setting ($5$, $0$, $-5$, $-10$, $-15$, and $-20\,$dB). 
To create these subsets, we mixed clean speech from \emph{LRS3} with noise files randomly sampled from the \emph{CHiME3} noise database~\cite{barker2015third}, adjusting the noise levels to achieve the specified \ac{SNR} for each set.

\subsection{Input representation}\vspace{-3px}
We use audio samples at a sampling rate of 16$\,$kHz. We encode the waveforms into mel-spectrograms with \ac{FFT} length of 1024, and hop length of 256 samples, thus obtaining a feature rate of 62.5$\,$Hz. The number of mel-frequency bins per frame is $n_a=80$.
The spectrograms are scaled and shifted to be standardized to zero-mean and a variance of 0.5, based on statistics computed on the training set. The time-frequency outputs of the model are converted into a time-domain signal using the neural vocoder $D_a$, for which we use the \emph{HiFi-GAN decoder} \cite{kong2020hifigan}, pretrained at 16$\,$kHz. 
Speaker embeddings are obtained via the \emph{Wespeaker} encoder $D_s$ \cite{wang2023wespeaker}, which encode speaker characteristics from an enrollment utterance of arbitrary length into a single feature vector of size $n_s=256$.
Video conditioning inputs are at 25 frames per second. 
The video frames are preprocessed to obtain the grayscale lip \ac{ROI}, following the pipeline in~\cite{martinez2020lipreading, ma2021end}.
The \acp{ROI} of 88 $\times$ 88 pixels are fed into the self-supervised model \emph{BRAVEn}~\cite{haliassos2024braven}, which results in video features of dimension $n_v=1024$ per video frame. 

\begin{table*}[t]
\centering
\caption{
Results on \emph{LRS3-CHiME3} showing mean values and standard deviation.
Comparison of speech enhancement  and lip-to-speech models at an \ac{SNR} of $-$10$\,$dB. 
Higher values indicate better performance except \ac{WER} and LSE-D. 
Input modalities include audio (A), video (V), enrollment audio (E), or their combinations. 
Enrollment audio is derived from the clean reference.
}
\label{tab:lrs3_results}
\resizebox{0.88\width}{0.88\height}{
\begin{tabular}{@{}c@{\hspace{0.7em}}lcccccccc}
\toprule
 &  & Input & DNSMOS $\uparrow$ & NISQA $\uparrow$ & LPS $\uparrow$ & WER $\downarrow$ & SpkSim $\uparrow$ & LSE-C $\uparrow$ & LSE-D$\downarrow$\\
\midrule
 & Clean & - & $3.56 \pm 0.36$ & $4.65 \pm 0.50$ & $1.00 \pm 0.00$ & $0.09 \pm 0.16$ & $1.00 \pm 0.00$ & $7.49 \pm 1.60$ & $7.10 \pm 0.96$ \\
 & Noisy & - & $2.16 \pm 0.14$ & $0.55 \pm 0.27$ & $0.11 \pm 0.19$ & $0.89 \pm 0.25$ & $0.26 \pm 0.24$ & $2.03 \pm 1.57$ & $11.15 \pm 1.53$ \\
\midrule
\multirow{2}{*}{\rotatebox{90}{\hspace{-0.2em}\scriptsize{SE}}}
 & \emph{SGMSE+} & A & $3.04 \pm 0.35$ & $2.73 \pm 0.87$ & $0.35 \pm 0.34$ & $0.75 \pm 0.37$ & $0.35 \pm 0.23$ & $4.79 \pm 2.01$ & $9.40 \pm 1.63$ \\
 & \emph{AV-Gen} & A+V & $2.90 \pm 0.25$ & $2.24 \pm 0.60$ & $0.50 \pm 0.33$ & $0.56 \pm 0.37$ & $0.36 \pm 0.22$ & $6.40 \pm 1.66$ & $7.76 \pm 1.06$ \\
\midrule
\multirow{6}{*}{\rotatebox{90}{\hspace{-0.2em}\scriptsize{Lip-to-Speech}}}
 & \emph{Lip2Speech} & V+E & $2.37 \pm 0.13$ & $1.42 \pm 0.43$ & $0.35 \pm 0.26$ & $0.72 \pm 0.29$ & $0.11 \pm 0.11$ & $4.29 \pm 2.43$ & $9.04 \pm 1.44$ \\
 & \emph{IL2S} & V+E & $2.84 \pm 0.26$ & $2.53 \pm 0.60$ & $0.52 \pm 0.27$ & $0.59 \pm 0.33$ & $0.34 \pm 0.14$ & $\mathbf{8.08 \pm 1.63}$ & $\mathbf{6.51 \pm 0.90}$ \\
 & \emph{IL2S} (SSL) & V+E & $2.86 \pm 0.28$ & $2.62 \pm 0.61$ & $\mathbf{0.67 \pm 0.28}$ & $0.36 \pm 0.34$ & $0.37 \pm 0.14$ & $8.03 \pm 1.61$ & $6.54 \pm 0.90$ \\
 & \emph{DiffV2S} & V & $3.17 \pm 0.29$ & $3.48 \pm 0.52$ & $0.45 \pm 0.37$ & $0.51 \pm 0.36$ & $0.16 \pm 0.13$ & $7.28 \pm 1.68$ & $7.27 \pm 1.02$ \\
 & \emph{LipVoicer} & V & $3.17 \pm 0.30$ & $3.53 \pm 0.64$ & $0.57 \pm 0.30$ & $\mathbf{0.36 \pm 0.33}$ & $0.15 \pm 0.14$ & $6.40 \pm 1.86$ & $8.12 \pm 1.22$ \\
 & \emph{LipDiffuser} (ours) & V+E & $\mathbf{3.64 \pm 0.30}$ & $\mathbf{4.57 \pm 0.52}$ & $0.64 \pm 0.27$ & $0.38 \pm 0.35$ & $\mathbf{0.63 \pm 0.14}$ & $6.84 \pm 1.60$ & $7.78 \pm 0.93$\\
\bottomrule
\end{tabular}
}
\end{table*}
\begin{figure}[t]
    \centering
    \begin{subfigure}[b]{0.40\textwidth}
        \includegraphics[width=\textwidth]{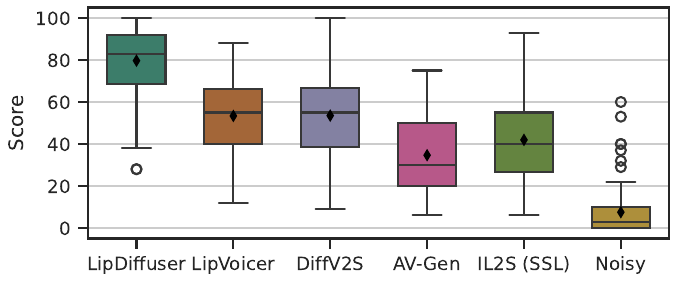}
        \caption{Speech quality}
        \label{fig:listening_qual}
    \end{subfigure}
    \hfill
    \begin{subfigure}[b]{0.40\textwidth}
        \includegraphics[width=\textwidth]{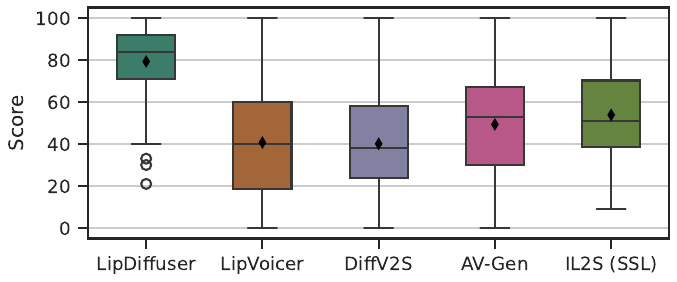}
        \caption{Speaker similarity}
        \label{fig:listening_sim}
    \end{subfigure}
\caption{Results of formal listening experiments. Participants were instructed to rate the overall speech quality or speaker similarity compared to a reference on a continuous scale from 0 to 100.} %
    \label{fig:listening_experiment}
\end{figure}

\subsection{Metrics} \vspace{-3px}
We evaluate the proposed \emph{LipDiffuser} method and compare it to other methods using a variety of metrics in order to capture different dimensions of performance.
To evaluate speech quality, we employ two non-intrusive, \ac{DNN}-based speech quality assessment models, \emph{DNSMOS}~\cite{reddy2021dnsmos} and \emph{NISQA}. 
 Moreover, we also evaluate the speaker similarity (SpkSim), measured as the cosine similarity between speaker embeddings generated from the enrollment and synthesized speech signals. %
To quantify the effect on \ac{ASR} tasks, we use the \texttt{Quartz\-Net\-15\-x5Base-En} model from the \emph{NeMo} toolkit \cite{kuchaiev2019nemo} as a downstream \ac{ASR} system and report the \acf{WER}. We also measure the \ac{LPS} using the \texttt{w2v LV-60K} \emph{wav2vec}-based phoneme predictor~\cite{pirklbauer2023evaluation}. 
To assess the temporal alignment between the generated speech and the corresponding lip movements, we employ automatic lip-sync metrics derived from \emph{SyncNet}~\cite{chung2017out}: \Ac{LSE-D} is the mean distance between feature embeddings of the video frames (focused on the mouth region) and their paired audio segment~\cite{prajwal2020lip}; \Ac{LSE-C} measures the average confidence with which SyncNet can associate each video-audio pair as being in sync~\cite{prajwal2020lip}.  

In addition to instrumental metrics, we also conducted a formal listening experiment with 15 participants, using randomly selected samples from the \emph{LRS3-CHiME3} dataset (at $-10$\,dB \ac{SNR}) and focusing on the aspects of speech quality and speaker similarity.%

\subsection{Baselines}\vspace{-3px}

For lip-to-speech methods, we compare with \emph{Lip2Speech}~\cite{kim2023lip}, \emph{DiffV2S}~\cite{choi2023diffv2s}, \emph{IL2S}~\cite{choi2023intelligible}, and \emph{LipVoicer}~\cite{yemini2024lipvoicer}, all of which were trained on \emph{LRS3}.
\emph{IL2S} comes in two configurations: the default with visual encoder trained from scratch, and a version which exploits visual features from a pretrained \ac{SSL} model.
As an audio-only \ac{SE} method, we utilize \emph{SGMSE+}~\cite{richter2023speech}.
We retrain the model with the original hyperparameters on the \emph{LRS3-CHiME3} dataset. As an \ac{AVSE} method, we employ \emph{AV-Gen}~\cite{richter2023audio}. 
This method integrates audio-visual features extracted from a pretrained \ac{SSL} model and inputs them into \emph{SGMSE+}~\cite{richter2023speech}. 
We retrain the model with the original hyperparameters using the \emph{LRS3-CHiME3} dataset.

\section{Results}\vspace{-5px}

\begin{figure}[t]
    \centering
        \includegraphics[width=0.7\linewidth]{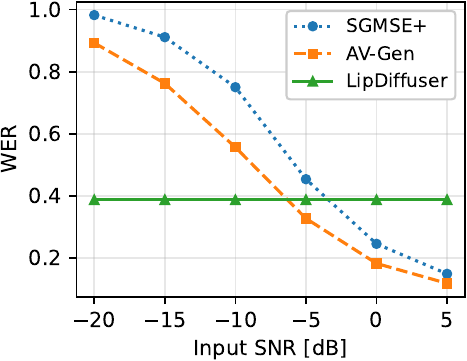}
\caption{ASR performance on the AVSE task in terms of \ac{WER}.}
    \label{fig:performance_over_dimensions}
\end{figure}

We begin with in-domain results using the \emph{LRS3-CHiME3} test set comprising 1321 test files. 
Table~\ref{tab:lrs3_results} shows the mean performance of \ac{SE} and lip-to-speech models at an \ac{SNR} of $-$10$\,$dB.
Among the \ac{SE} methods, the audio-visual \emph{AV-Gen} model outperforms the audio-only \emph{SGMSE+} baseline regarding WER, and LPS, highlighting the benefit of visual information. 
For lip-to-speech approaches, our proposed \emph{LipDiffuser} achieves the highest speech quality scores 
and the strongest speaker similarity,
approaching the clean reference. 
In terms of \ac{ASR} performance, \emph{LipDiffuser}, \emph{LipVoicer}, and \emph{IL2S (SSL)} achieve comparable \ac{WER} and LPS, all substantially outperforming the \ac{SE} methods. 
The results of our listening experiments, depicted in Figure~\ref{fig:listening_experiment}, follow a similar trend to the instrumental metrics w.r.t. speech quality and speaker similarity, with \emph{LipDiffuser} showing a high mean score compared to the baselines, as well as a relatively narrower score distribution for both speech quality and speaker similarity.

Figure~\ref{fig:performance_over_dimensions} shows WER for \emph{SGMSE+}, \emph{AV-Gen}, and \emph{LipDiffuser} as a function of input \ac{SNR}. 
While the performance of both \ac{SE} models declines with decreasing SNR, \emph{LipDiffuser's} performance remains constant since it does not rely on input audio. 
We observe that lip-to-speech consistently outperforms \ac{SE} below the $-$5$\,$dB \ac{SNR} threshold, demonstrating its superiority in severely noisy environments.

\section{Conclusion}\vspace{-5px}
In this work, we proposed \emph{LipDiffuser}, a lip-to-speech diffusion model that synthesizes high-quality and intelligible speech from silent video recordings and an enrollment utterance. 
We proposed magnitude-preserving feature-wise linear modulation (MP-FiLM) layers for feature fusion within the magnitude-preserving ablated diffusion model (MP-ADM) network architecture. 
Experiments show that below the threshold of $-$5$\,$dB input \ac{SNR}, \emph{LipDiffuser} outperforms audio-visual and audio-only \ac{SE} baselines
and consistently outperforms lip-to-speech baselines in audio quality and speaker similarity, with some robustness w.r.t different enrollment utterance scenarios. 
\emph{LipDiffuser} demonstrates the strong capabilities of conditional diffusion models for generating high-quality speech in the lip-to-speech task.

\section{References}\vspace{-5px}
\label{sec:refs}
\atColsBreak{\vskip5pt}
\printbibliography[heading=none]

\end{document}